\documentclass[10pt]{amsart}
\usepackage{graphicx}
\usepackage{amsmath}
\usepackage{amsthm}
\usepackage{amsfonts}
\usepackage{amssymb}
\usepackage{cite}

\usepackage{epic}
\usepackage{pstricks}

\theoremstyle{plain}
\newtheorem{theorem}{Theorem}[section]
\newtheorem{proposition}[theorem]{Proposition}

\newtheorem{lemma}[theorem]{Lemma}
\newtheorem{sublemma}[theorem]{Sublemma}

\newtheorem{definition}[theorem]{Definition}
\newtheorem{remark}[theorem]{Remark}

\newcommand{\lp}{\left(}
\newcommand{\rp}{\right)}

\newcommand{\nn}{\overrightarrow{n}}

\renewcommand{\v}[2]{\overrightarrow{V_{#1}V_{#2}}}

\newcommand{\intr}[2]{\overline{#1,#2}}

\renewcommand{\c}[2]{$\mathrm{C}_{#1}\!\lp#2\rp$}
\newcommand{\M}[2]{$\mathrm{M}_{#1}(#2)$}

\newcommand{\nega}{\overset{\displaystyle{\neg}}{\rule{0pt}{1.5pt}}}

\renewcommand{\|}{\,\rule[-3.5pt]{.5pt}{12pt}\,}
\newcommand{\s}{\,\rule[-3.5pt]{1pt}{12pt}\,}

\newcommand{\De}{\Delta}
\newcommand{\al}{\alpha}
\newcommand{\be}{\beta}
\newcommand{\la}{\lambda}

\newcommand{\g}{\gamma}
\newcommand{\vp}{\varepsilon}

\renewcommand{\le}{\leqslant}
\renewcommand{\ge}{\geqslant}

\newcommand{\R}{\mathbb{R}}
\newcommand{\Z}{\mathbb{Z}}

\renewcommand{\P}{\mathcal{P}}

\newcommand{\conv}{\operatorname{conv}}

\newcommand{\sign}{\operatorname{sign}}

\begin{document}


\title[An $O(n)$ Polygon Convexity Test]
{Polygon Convexity: 
A Minimal $O(n)$ Test} 


\author{Iosif Pinelis}  
                   

\date{\today; file convex-poly/test/Sep06/\jobname}

\begin{abstract}
An $O(n)$ test for polygon convexity is stated and proved. It is also proved that the test is minimal in a certain exact sense.
\end{abstract}



\maketitle

\date{\today. File: { ipinelis/polygon/convex-poly/sub-polygons/\jobname.tex
}
}

\setcounter{section}{-1}

\section{Introduction}



Everyone knows a convex polygon when one sees it. Yet, to deal with the notion of polygon convexity mathematically or computationally, it must be adequately described.  
A convex polygon can be defined, as e.g. in \cite[page 5]{yaglom}, as 
a succession of connected line segments which constitute the boundary of a convex set.
However, in computational geometry it seems more convenient to
consider a polygon as a sequence of its vertices, say $(V_0,\dots,V_{n-1})$, with the edges being the segments $[V_0,V_1],\dots,[V_{n-2},V_{n-1}],[V_{n-1},V_0]$. 
Then one can say that a polygon is convex if the union of its edges coincides with the boundary of the convex hull of the set of vertices $\{V_0,\dots,V_{n-1}\}$.
 
One finds the following statement in \cite[page 233]{moret}:
\begin{quote}
{\bf Theorem 4.3}\quad Let the sequence of vertices, $p_1,p_2,\dots,p_n,p_{n+1}=p_1$, define an arbitrary polygon $P$ and let $P_i$ be the polygon defined by the sequence of vertices $p_1,p_2,\dots,p_i,p_1$. Then $P$ is convex if and only if, for each $i$,\quad $i=3,4,\dots,n$, polygon $P_i$ is itself convex. 
\end{quote}
It is also said in \cite{moret} that an
incremental test for polygon convexity can be based on the quoted theorem. No proof or reference to a proof of this theorem was given there. Moreover, the ``if" part of the theorem is trivial: if all polygons $P_3,\dots,P_n$ are convex, then polygon $P=P_n$ is trivially convex. Thus, such a theorem by itself
would be impossible to use for an incremental test. 

One might suppose that there was a typo in the quoted statement of Theorem~4.3 and there was meant to be $i=3,4,\dots,n-1$ in place of $i=3,4,\dots,n$ (or, equivalently, $p_1,p_2,\dots,p_n,p_{n+1},p_1$ in place of $p_1,p_2,\dots,p_n,p_{n+1}=p_1$). 
But then
the theorem could not be true. 
Indeed, note that all $n$-gons with $n\le3$ are convex. Hence, if the ``if" part of quoted Theorem 4.3 were true with $i=3,4,\dots,n-1$ in place of $i=3,4,\dots,n$, then it would immediately follow by induction in $n$ that all polygons whatsoever are convex! 

However, it appears that the polygon convexity test suggested in \cite{moret} may be basically correct by itself, even though it is not in fact based on the quoted theorem (or proved otherwise). 
In this paper, we rigorously state and prove 
an $O(n)$ polygon convexity test, which is similar to the test suggested in \cite{moret}. 
Moreover, we show that our test is minimal in the sense that none of the $3(n-3)$ test conditions can be dropped if the test is to remain valid. 

Under the additional condition that the $n$-gon is simple (that is, the only points belonging to two different edges of the $n$-gon are its vertices), an $O(n)$ convexity test seems to be well known \cite{moret,hill,weisstein} but hardly ever rigorously proved. However, no $O(n)$ simplicity tests seem to be known \cite{moret}. 

One may also note that the ``only if" part of the quoted Theorem 4.3 turns out basically correct. Indeed, the main result in \cite{elimin} states that 
if $\P=(V_0,\dots,V_{n-1})$ is
a convex polygon whose vertices are all {\em distinct}, then the reduced polygon $\P^{(i)}:=(V_0,\dots,V_{i-1},V_{i+1},\dots,V_{n-1})$ (with vertex $V_i$ and hence edges $[V_{i-1},V_i]$ and $[V_i,V_{i+1}]$ removed) is also convex, for each $i$.  

In addition to such downward hereditariness of polygon convexity, it is shown in \cite{elimin} that the polygon convexity property is hereditary upwards as well. Namely, if a polygon $\P=(V_0,\dots,V_{n-1})$ with $n\ge5$ vertices is such that all the reduced polygons $\P^{(i)}$ are convex, then
$\P$ is also convex.

Taken together, the downward and upward hereditariness of polygon convexity can be used to obtain conditions necessary and sufficient for polygon convexity. In particular, a corollary in \cite{elimin} states that a polygon $\P=(V_0,\dots,V_{n-1})$ with $n\ge5$ distinct vertices is convex if and only if all the reduced polygons $\P^{(i)}$ are convex. 
Such a test is helpful in theoretical considerations. However, it would be extremely wasteful computationally, as it
takes $\Omega(n!)$ operations.

The paper is organized as follows. 
In Section \ref{results}, the basic definitions are given and the main results are stated: Theorem~\ref{th:calculation}, which provides an $O(n)$ polygon convexity test; and Proposition \ref{prop:minimality}, which shows that the test is exactly minimal in a certain sense.

In Section \ref{proofs}, the proofs are given. 
More specifically, Subsection \ref{subsec:defs} of Section~\ref{proofs} contains definitions needed in the proofs. 
Subsection~\ref{subsec:lemmas} contains
statements of lemmas and based on them proofs of the main results stated in Section \ref{results}; the proofs of all lemmas are deferred further to Subsection \ref{subsec:proofs of lemmas}.  

\section{Definitions and results}\label{results}

A {\em polygon} is defined in this paper as any finite sequence of points (or, interchangeably, vectors) on the Euclidean plane $\R^2$. Let here $\P:=(V_0,\dots,V_{n-1})$ be a polygon, which is sequence of $n$ points; such a polygon is also called an $n$-gon. 
The points $V_0,\dots,V_{n-1}$ are called the {\em vertices} of $\P$.
The smallest value that one may allow for the integer $n$ is $0$, corresponding to a polygon with no vertices, that is, to the sequence $()$ of length $0$. 
The segments, or closed intervals,
$$[V_i,V_{i+1}]:=\conv\{V_i,V_{i+1}\}\quad\text{for}\ i\in\{0,\dots,n-1\}$$
are called the {\em edges} of polygon $\P$, where 
$$V_n:=V_0.$$
The symbol $\conv$ denotes, as usual, the convex hull \cite[page 12]{rock}.
Note that, if $V_i=V_{i+1}$, then the edge $[V_i,V_{i+1}]$ is a singleton set. 
 
In general, our terminology corresponds to that in \cite{rock}. 
Here and in the sequel, we also use the notation
$$\intr km:=\{i\in\Z\colon k\le i\le m\},$$
where $\Z$ is the set of all integers; in particular, $\intr km$ is empty if $m<k$. 

Let us define the convex hull and dimension of polygon $\P$ as, respectively, the convex hull and dimension of the set of its vertices: $\conv\P:=\conv\{V_0,\dots,V_{n-1}\}$ and $\dim\P:=\dim\{V_0,\dots,V_{n-1}\}=\dim\conv\P$.


Given the above notion of the polygon, a {\em convex polygon} can be defined as a polygon $\P$ such that the union of the edges of $\P$ coincides with the boundary $\partial\conv\P$ of the convex hull $\conv\P$ of $\P$; cf. e.g. \cite[page 5]{yaglom}. 
Thus, one has

\begin{definition}\label{def:conv}
A polygon $\P=(V_0,\dots,V_{n-1})$ is {\em convex} if 
$$\bigcup_{i\in\intr0{n-1}}[V_i,V_{i+1}]=\partial\conv\P.$$
\end{definition}

Let us emphasize that a polygon in this paper is a sequence and therefore ordered. In particular, even if all the vertices $V_0,\dots,V_{n-1}$ of a polygon $\P=(V_0,\dots,V_{n-1})$ are the extreme points of the convex hull of $\P$, it does not necessarily follow that $\P$ is convex. For example, consider the points $V_0=(0,0)$, $V_1=(1,0)$, $V_2=(1,1)$, and $V_3=(0,1)$. Then polygon $(V_0,V_1,V_2,V_3)$ is convex, while polygon $(V_0,V_2,V_1,V_3)$ is not.

In this paper, we shall be concerned foremost with strict convexity. 

\begin{definition}\label{def:strict}
Let us say that a polygon $\P=(V_0,\dots,V_{n-1})$ is {\em strict} if 
for any three distinct $i$, $j$, and $k$ in the set $\intr0{n-1}$, the vertices $V_i$, $V_j$, and $V_k$ are non-collinear. 
\end{definition}

\begin{definition}\label{def:strict conv}
Let us say that a polygon is {\em strictly convex} if 
it is both strict and convex. 
\end{definition}

\begin{remark}\label{rem:test}
Any $3$-gon is convex, and so, a $3$-gon is strictly convex if and only if it is strict. All $n$-gons with $n\le2$ are strictly convex.
\end{remark}

For a polygon $\P=(V_0,\dots,V_{n-1})$, let $x_i$ and $y_i$ denote the coordinates of its vertices $V_i$, so that
$$V_i=(x_i,y_i)\quad\text{for}\ i\in\intr0{n-1}.$$ 
Introduce the determinants
\begin{equation}\label{eq:De}
\De_{\al,i,j}:=
\left|
\begin{matrix}
1&x_\al&y_\al \\
1&x_i&y_i \\
1&x_j&y_j 
\end{matrix}\right|
\end{equation}
for $\al$, $i$, and $j$ in the set $\intr0{n-1}$. 
Let then
$$
\begin{aligned}
a_i&:=\sign\De_{i+1,i-1,i}=\sign\De_{i-1,i,i+1};\\
b_i&:=\sign\De_{0,i-1,i};\\
c_i&:=\sign\De_{i,0,1}=\sign\De_{0,1,i}.
\end{aligned}
$$

The following theorem is the main result of this paper, which provides an $O(n)$ test of the strict convexity of a polygon.

\begin{theorem}\label{th:calculation}
An $n$-gon $\P=(V_0,\dots,V_{n-1})$ with $n\ge4$ is strictly convex if and only if conditions
\begin{equation}\label{eq:conds}
	\begin{aligned}
a_i b_i& >0,\\
a_i b_{i+1}& >0,\\
c_i c_{i+1}& >0
\end{aligned}
\end{equation}
hold for all
$$i\in\intr2{n-2}.$$
\end{theorem}

\begin{proposition}\label{prop:minimality}
None of the $3(n-3)$ conditions in Theorem~\ref{th:calculation} can be omitted without (the ``if" part of) Theorem~\ref{th:calculation} ceasing to hold. 
\end{proposition}

Thus, the test given by Theorem~\ref{th:calculation} is exactly minimal.

\begin{remark}\label{rem:minimality}
Adding to the $3(n-3)$ conditions \eqref{eq:conds} in Theorem~\ref{th:calculation} the equality $b_2=c_2$, which trivially holds for any polygon (convex or not), one can rewrite \eqref{eq:conds} as the following system of $3(n-3)+1$ equations and one inequality:
$$
	\begin{aligned}
&a_2=\dots=a_{n-2}\\
=&b_2=\dots=b_{n-2}=b_{n-1}\\
=&c_2=\dots=c_{n-2}=c_{n-1}\ne0.
\end{aligned}
$$
\end{remark}

\section{Proofs}\label{proofs}

\subsection{More Definitions}\label{subsec:defs}

\begin{definition}\label{def:ordinary}
Let us say that a polygon $\P=(V_0,\dots,V_{n-1})$ is {\em ordinary} if 
its vertices are all distinct from one another: $(i\ne j\ \&\ i\in\intr0{n-1}\ \&\ j\in\intr0{n-1}\,)\implies V_i\ne V_j$.
\end{definition}

Let us say that two vertices of a polygon $\P=(V_0,\dots,V_{n-1})$ are {\em adjacent} if they are the two endpoints of an edge of $\P$; thus, 
$$\{V_0,V_1\},\{V_1,V_2\},\dots,\{V_{n-2},V_{n-1}\},\{V_{n-1},V_0\}$$
are the pairs of adjacent vertices of polygon $\P=(V_0,\dots,V_{n-1})$. 

\begin{definition}\label{def:quasi-strict}
Let us say that a polygon $\P$
is {\em quasi-strict} if
any two adjacent vertices of $\P$ are not collinear with any other vertex of $\P$. More formally, a polygon $\P$
is {\em quasi-strict} if,
for any $i\in\intr0{n-1}$ and any $j\in\intr0{n-1}\setminus\{i,i\oplus1\}$, the points $V_i$, $V_{i\oplus1}$, and $V_j$ are non-collinear, where 
$$i\oplus1:=
\begin{cases}
i+1 &\text{if}\ i\in\intr0{n-2},\\
0 &\text{if}\ i=n-1. 
\end{cases}
$$ 
\end{definition}

\begin{definition}\label{def:quasi-strictly-convex}
Let us say that a polygon 
is {\em quasi-strictly convex} if
it is both convex and quasi-strict. 
\end{definition}

\begin{definition}\label{def:to one side}
Let $P_0,\dots,P_m$ be any points on the plane, any two of which may in general coincide with each other. Let us write $P_2,\dots,P_m\|[P_0,P_1]$ 
and say
that
points $P_2,\dots,P_m$ are {\em to one side} 
of segment $[P_0,P_1]$ if 
there is a (straight) line $\ell$ containing $[P_0,P_1]$ and supporting to the set $\{P_0,\dots,P_m\}$; the latter, ``supporting" condition means here (in accordance with \cite[page 100]{rock}) that $\ell$ is the boundary of a closed 
half-plane containing the set $\{P_0,\dots,P_m\}$. 

Let us write $P_2,\dots,P_m\s[P_0,P_1]$ 
and say
that
points $P_2,\dots,P_m$ are {\em strictly to one side}) 
of segment $[P_0,P_1]$ if $P_2,\dots,P_m\|[P_0,P_1]$ and none of the points $P_2,\dots,P_m$ is collinear with points $P_0$ and $P_1$.

For any given $i\in\intr0{n-1}$, let us say that
a polygon $\P=(V_0,\dots,V_{n-1})$ is to one side 
(respectively, strictly to one side) 
of its edge $[V_i,V_{i+1}]$ if the points of the set 
$\{V_j\colon j\in\intr0{n-1}\setminus\{i,i\oplus1\}\}$ 
are so.

Let us say that a polygon is {\em (strictly) to-one-side} if it is (strictly) to one side of every one of its edges. 
\end{definition}

\subsection{Lemmas, and Proofs of Theorem \ref{th:calculation} and Proposition \ref{prop:minimality}} \label{subsec:lemmas}

\begin{lemma}\label{lem:ordinary}
If an $n$-gon $\P=(V_0,\dots,V_{n-1})$ with $n\ge3$ is quasi-strict, then it is ordinary.
\end{lemma}

\begin{lemma}\label{lem:def}
An $n$-gon with $n\ge3$ is quasi-strictly convex if and only if 
it is strictly to-one-side. 
\end{lemma}

\begin{lemma}\label{lem:calculation}
Let $x_i$ and $y_i$ denote the coordinates of points $V_i$, so that
$V_i=(x_i,y_i)$ for all $i\in\intr0{n-1}$.
Then, for any choice of $\al$, $\be$, $i$, and $j$ in $\intr0{n-1}$, 
$$V_\al,V_\be\s[V_i,V_j]\iff
\De_{\al,i,j}\,\De_{\be,i,j}>0,$$
where $\De_{\al,i,j}$ are given by \eqref{eq:De}.
\end{lemma}

\begin{lemma}\label{lem:test}
For any $n\ge4$, a polygon $\P=(V_0,\dots,V_{n-1})$ is quasi-strictly convex if and only if conditions
\begin{align}
V_{i+1},V_0 & \s[V_{i-1},V_i],\label{eq:P1i}\tag{\c1 i} \\
V_{i-1},V_0 & \s[V_i,V_{i+1}],\label{eq:P2i}\tag{\c2 i} \\
V_i,V_{i+1} & \s[V_0,V_1].\label{eq:P3i}\tag{\c3 i}
\end{align}
hold for all
$$i\in\intr2{n-2}.$$
\end{lemma}

\begin{lemma}\label{lem:minimality}
None of the $3(n-3)$ conditions \emph{(\c\omega i)}
($\omega\in\{1,2,3\}$, $i\in\intr2{n-2}$) in Lemma \ref{lem:test} can be omitted without (the ``if" part of) Lemma \ref{lem:test} ceasing to hold. 
\end{lemma}

\begin{lemma}\label{lem:elim}
If a polygon $\P=(V_0,\dots,V_{n-1})$ is quasi-strictly convex, then it remains so after the elimination of any one (and hence any number) of its vertices; in particular, then the polygon $\P_{n-1}:=(V_0,\dots,V_{n-2})$ is quasi-strictly convex.
\end{lemma}

(Cf. the main result in \cite{elimin}.)

\begin{lemma}\label{lem:strict}
A convex polygon is strict if and only if 
it is quasi-strict. 
\end{lemma}

\begin{proof}[Proof of Theorem \ref{th:calculation}]
This follows immediately from Lemma \ref{lem:strict}, Lemma \ref{lem:test}, and Lemma \ref{lem:calculation}.
\end{proof}

\begin{proof}[Proof of Proposition \ref{prop:minimality}]
This follows immediately from Lemma \ref{lem:strict}, Lemma \ref{lem:minimality}, and Lemma \ref{lem:calculation}.
\end{proof}

\subsection{Proofs of the Lemmas}\label{subsec:proofs of lemmas}

\begin{proof}[Proof of Lemma \ref{lem:ordinary}]
Indeed, if $V_i=V_j$ while $0\le i<j\le n-1$, then (recalling Definition \ref{def:quasi-strict}) one sees that $i\oplus1=i+1$ and the points $V_i$, $V_{i\oplus1}$, and $V_j$ are collinear. 

If at that $j\ne i+1$, then $j\in\intr0{n-1}\setminus\{i,i\oplus1\}$, so that polygon $\P=(V_0,\dots,V_{n-1})$ is not quasi-strict. 
Next, 
the set $\intr0{n-1}\setminus\{i,i\oplus1\}$ is non-empty (because $n\ge3$), so that there exists some $k\in\intr0{n-1}\setminus\{i,i\oplus1\}$.
If now 
$j=i+1$, then 
the three points $V_i$, $V_{i\oplus1}=V_{i+1}=V_j=V_i$, and $V_k$ are trivially collinear, so that again one concludes that $\P$ is not quasi-strict. 
\end{proof}


\begin{proof}[Proof of Lemma \ref{lem:def}]
Observe first that a polygon is strictly to-one-side if and only if it is quasi-strict and to-one-side. (This follows immediately from Definitions \ref{def:to one side} and \ref{def:quasi-strict}.) Also, by Lemma \ref{lem:ordinary}, every quasi-strict $n$-gon with $n\ge3$ is ordinary. On the other hand, it was shown in \cite
{elimin} that an ordinary polygon is convex if and only if it is to-one-side. Now Lemma \ref{lem:def} follows.
\end{proof}

\begin{proof}[Proof of Lemma \ref{lem:calculation}]
Take any $\al$, $\be$, $i$, $j$ in the set $\intr0{n-1}$. By Definition \ref{def:to one side}, one has $V_\al,V_\be\s[V_i,V_j]$ if and only if $V_j\ne V_i$ and there exists some vector $\nn=(a,b)\in\R^2$ such that
$$\nn\cdot\v ij=0<\nn\cdot\v i\g\quad\text{for}\ \g\in\{\al,\be\}.$$
Since
$$\De_{\al,i,j}=
\left|
\begin{matrix}
1&x_\al-x_i&y_\al-y_i \\
1&0&0 \\
1&x_j-x_i&y_j-y_i 
\end{matrix}\right|,
$$
one may replace without loss of generality (w.l.o.g.) the points $V_\al$, $V_\be$, $V_i$, $V_j$ by $V_\al-V_i$, $V_\be-V_i$, $V_i-V_i=(0,0)$, $V_j-V_i$, respectively. Hence, w.l.o.g.\  
$$V_i=(0,0).$$
Then the condition $V_\al,V_\be\s[V_i,V_j]$ can be rewritten as follows:
$$(x_j,y_j)\ne(0,0)\quad\text{and}\quad ax_j+by_j=0<ax_\g+by_\g\quad\text{for}\ \g\in\{\al,\be\}.$$
W.l.o.g., $y_j\ne0$. 
Then condition $ax_j+by_j=0$ is equivalent to $b=-\frac{x_j}{y_j}a$, so that the inequality $0<ax_\g+by_\g$ can be rewritten as 
$\frac a{y_j}(x_\g y_j-x_j y_\g)>0$, or as $\frac a{y_j}\De_{\g,i,j}<0$ (where $\g\in\{\al,\be\}$); in particular, it follows that $a\ne0$.

We see that the condition $V_\al,V_\be\s[V_i,V_j]$ implies 
$$\De_{\al,i,j}\De_{\be,i,j}
=\lp\frac{y_j}a\rp^2 \lp\frac a{y_j}\De_{\al,i,j}\rp\lp\frac a{y_j}\De_{\be,i,j}\rp
>0.$$
This proves the ``$\Longrightarrow$" part of Lemma \ref{lem:calculation}. 

To prove the ``$\Longleftarrow$" part, let $\nn:=\vp(-y_j,x_j)$, where $\vp:=\sign\De_{\al,i,j}$. Then the condition $\De_{\al,i,j}\De_{\be,i,j}>0$ implies that $\vp=\sign\De_{\be,i,j}$. Also, one has $\nn\cdot\v ij=0$, while
$$\nn\cdot\v i\g=\vp(x_j y_\g-y_j x_\g)=\vp\De_{\g,i,j}=|\De_{\g,i,j}|>0$$
for $\g\in\{\al,\be\}$, so that the condition $V_\al,V_\be\s[V_i,V_j]$ takes place.
\end{proof}

\begin{proof}[Proof of Lemma \ref{lem:test}]

\textbf{``Only if"}\quad This part of Lemma \ref{lem:test} follows immediately from Lemma \ref{lem:def}. 

\textbf{``If"}\quad Assume that indeed conditions (\c1 i), (\c2 i), and (\c3 i) hold for all $i\in\intr2{n-2}$. To prove the ``if" part of Lemma \ref{lem:test}, it suffices to show that, for for all $k\in\intr3n$, the polygon $\P_k:=(V_0,\dots,V_{k-1})$ is quasi-strictly convex. We shall do this by induction in $k$. 

For $k=3$, the polygon $\P_k=\P_3=(V_0,V_1,V_2)$ is quasi-strict, in view of condition (\c1{2}) and Definitions \ref{def:to one side} and \ref{def:quasi-strict}. Therefore, $\P_k$ is quasi-strictly convex for $k=3$. 

Suppose now that 
$$k\in\intr3{n-1}$$
and $\P_k$ is quasi-strictly convex. We have then to verify that polygon $\P_{k+1}=(V_0,\dots,V_k)$ is quasi-strictly convex. 

Since $k\in\intr3{n-1}$, one has $k-1\in\intr2{n-2}$.
Hence, 
condition (\c2{k-1}) holds, and it implies that the points $V_0$, $V_{k-1}$, and $V_k$ are non-collinear. Therefore, w.l.o.g.\ 
$$V_k=(0,0),\quad V_0=(1,0),\quad V_{k-1}=(0,1).$$
Let also
$$V_{k-2}=(u,v)\quad\text{and}\quad V_1=(x,y),$$
for some real $x$, $y$, $u$, and $v$. 
Finally, take any 
$$i\in\intr1{k-2}$$
and let
$$V_i=(\la,\mu),$$
for some real $\la$ and $\mu$. 

Since $k-1\in\intr2{n-2}$, 
conditions (\c1{k-1}), (\c2{k-1}), (\c3{k-1}) hold. In view of Lemma \ref{lem:calculation}, these three conditions yield respectively that
\begin{gather}
u\,(u+v-1)>0, \label{eq:P1(k-1)} \\
u>0, \label{eq:P2(k-1)} \\
(x+y-1)\,y>0. \label{eq:P3(k-1)}
\end{gather}

Because polygon $\P_k=(V_0,\dots,V_{k-1})$ is assumed to be quasi-strictly convex, it follows by Lemma \ref{lem:def} that $\P_k$ is strictly to one side of every one of its edges, 
$$[V_0,V_1],\dots,[V_{k-2},V_{k-1}],[V_{k-1},V_0].$$
In particular, one has $V_{k-2},V_1\s[V_0,V_{k-1}]$ (because the condition $k\in\intr3{n-1}$ implies that $k-2\ne0$ and $1\ne k-1$). 
In view of Lemma \ref{lem:calculation}, this yields 
$$(1-u-v)(1-x-y)>0.$$
Now it follows from \eqref{eq:P1(k-1)}--\eqref{eq:P3(k-1)} that
\begin{align}
u & >0, \label{eq:1} \\
u+v-1 & >0, \label{eq:2} \\
x+y-1 & >0, \label{eq:3} \\
y & >0. \label{eq:4} 
\end{align}
Moreover, 
the quasi-strict convexity of polygon $\P_k$ and Lemma \ref{lem:def} imply relations
$$V_i,V_1\s[V_0,V_{k-1}],\quad V_i,V_{k-1}\|[V_0,V_1],\quad V_i,V_0\|[V_{k-2},V_{k-1}],$$
which in turn yield 
\begin{align*}
(1-\mu-\la)(1-x-y)&>0,\\
((1-\la)y+\mu(x-1))(x+y-1)&\ge0,\\
((1-\mu)u+\la(v-1))(u+v-1)&\ge0,
\end{align*}
respectively (the last two inequalities are in fact strict except for the cases $i=1$ for the former and $i=k-2$ for the latter). 
In view of \eqref{eq:3} and \eqref{eq:2}, these three inequalities imply
\begin{align}
\la+\mu-1 & >0, \label{eq:8} \\
(1-\la)y+\mu(x-1) & \ge0, \label{eq:6} \\
(1-\mu)u+\la(v-1) & \ge0. \label{eq:7} 
\end{align}

Next, 
\eqref{eq:1} and \eqref{eq:8} imply $u\,(\la+\mu-1)>0$. Adding this inequality to \eqref{eq:7}, one has $\la\,(u+v-1)>0$. Now \eqref{eq:2} yields
\begin{equation}\label{eq:10} 
\la>0.
\end{equation}
In view of Lemma \ref{lem:calculation}, this is equivalent to
$V_i,V_0\s[V_{k-1},V_k]$, for all $i\in\intr1{k-2}$. That is,
\begin{equation}\label{eq:10a} 
V_0,\dots,V_{k-2}\s[V_{k-1},V_k].
\end{equation} 

Similarly, \eqref{eq:4} and \eqref{eq:8} imply $y\,(\la+\mu-1)>0$. Adding this inequality to \eqref{eq:6}, one has $\mu\,(x+y-1)>0$. Now \eqref{eq:3} yields
\begin{equation}\label{eq:9} 
\mu>0,
\end{equation}
which is equivalent to
$V_i,V_{k-1}\s[V_k,V_0]$, for all $i\in\intr1{k-2}$. That is,
\begin{equation}\label{eq:9a} 
V_1,\dots,V_{k-1}\s[V_k,V_0].
\end{equation} 

Also, since condition (\c3 i) was assumed to hold for all $i\in\intr2{n-2}$, one has $V_2,\dots,V_{n-1}\s[V_0,V_1]$. Hence and because $k\in\intr3{n-1}$, 
\begin{equation}\label{eq:P3} 
V_2,\dots,V_k\s[V_0,V_1].
\end{equation} 

Suppose that the following sublemma of Lemma \ref{lem:test} is true (we shall prove the sublemma after the proof of Lemma \ref{lem:test} is completed). 

\begin{sublemma} \label{sublem:test}
For all $i\in\intr1{k-2}$, 
one has $V_k,V_0\s[V_i,V_{i+1}]$.
\end{sublemma}

Let us now complete the proof of Lemma \ref{lem:test}. Since polygon $\P_k$ is assumed to be quasi-strictly convex, Lemma \ref{lem:def} implies that, for all $i\in\intr1{k-2}$, $$V_0,\dots,V_{i-1},V_{i+2},\dots,V_{k-1}\s[V_i,V_{i+1}],$$
and so, by Sublemma \ref{sublem:test},
\begin{equation}\label{eq:test} 
V_0,\dots,V_{i-1},V_{i+2},\dots,V_{k-1},V_k\s[V_i,V_{i+1}]\quad \text{for all }
i\in\intr1{k-2}.
\end{equation} 
Relations \eqref{eq:10a}, \eqref{eq:9a}, \eqref{eq:P3}, and \eqref{eq:test} taken together mean that polygon $\P_k$ is strictly to one side of every one of its edges, 
$$[V_0,V_1],\dots,[V_{k-1},V_k],[V_k,V_0].$$ 
Hence, by Lemma \ref{lem:def}, polygon $\P_{k+1}$ is quasi-strictly convex. Thus, the induction step is verified. 
\end{proof}

\begin{proof}[Proof of Sublemma \ref{sublem:test}]
Take any $i\in\intr1{k-2}$ and let
$$V_{i+1}=(a,b),$$
for some real $a$ and $b$. We need to show that $V_k,V_0\s[V_i,V_{i+1}]$. 
If $i=k-2$, then condition $V_k,V_0\s[V_i,V_{i+1}]$ coincides with condition (\c1{k-1}). Hence, w.l.o.g.\ 
$$i\in\intr1{k-3},$$
so that $\{k-1,0\}\cap\{i,i+1\}=\emptyset$.
Therefore and because polygon $\P_k=(V_0,\dots,V_{k-1})$ was assumed to be quasi-strictly convex, Lemma \ref{lem:def} yields $V_{k-1},V_0\s[V_i,V_{i+1}]$. By Lemma \ref{lem:calculation}, the latter relation can be rewritten as
\begin{equation}\label{eq:11} 
(\la q-\mu p+p)(\la q-\mu p-q)>0,
\end{equation} 
where 
$$p:=a-\la\quad\text{and}\quad q:=b-\mu.$$
 
On the other hand, relation $V_k,V_0\s[V_i,V_{i+1}]$ (which is to be proved here) can be rewritten as
\begin{equation}\label{eq:12} 
(\la q-\mu p)(\la q-\mu p-q)>0.
\end{equation} 

Consider separately the following three cases, depending on whether $\la q-\mu p$ is zero, positive, or negative.

{\em Case 1\/}: $\la q-\mu p=0$.\quad Then \eqref{eq:10} and \eqref{eq:9} yield $pq\ge0$, while \eqref{eq:11} implies $pq<0$, which is a contradiction.

{\em Case 2\/}: $\la q-\mu p>0$.\quad Here, if \eqref{eq:12} failed to hold, one would have 
\begin{equation}\label{eq:13} 
\la q-\mu p-q\le0
\end{equation}
and hence also  
\begin{equation}\label{eq:13a} 
q>0.
\end{equation}
Now \eqref{eq:11} would imply
\begin{equation}\label{eq:14} 
\la q-\mu p+p<0
\end{equation}
and hence also  
\begin{equation}\label{eq:14a} 
p<0.
\end{equation}
Next, \eqref{eq:13a} and \eqref{eq:8} would yield $(-q)(\la+\mu-1)<0$. Adding the latter inequality to \eqref{eq:13}, one would have $(-\mu)(p+q)<0$, which   would result (in view of \eqref{eq:9}) in
\begin{equation}\label{eq:15} 
p+q>0.
\end{equation}
On the other hand, \eqref{eq:14a} and \eqref{eq:8} would yield $p\,(\la+\mu-1)<0$. Adding the latter inequality to \eqref{eq:14}, one would have $\la\,(p+q)<0$ and then, in view of \eqref{eq:10},
$p+q<0$, which would contradict \eqref{eq:15}. 

{\em Case 3\/}: $\la q-\mu p<0$.\quad This case is quite similar to Case 2: just switch the direction of all inequalities obtained in the consideration of Case 2.
\end{proof}

\begin{proof}[Proof of Lemma \ref{lem:minimality}]
The proof is based on  

\begin{sublemma}\label{lem:minimal}
Let $\P_k:=(V_0,\dots,V_{k-1})$ be any quasi-strict $k$-gon with $k\ge3$.
Then
\begin{description}
\item[(i)] there exists a point $V_k$ such that 
the $(k+1)$-gon $\P_{k+1}:=(V_0,\dots,V_{k-1},V_k)$ is quasi-strict and satisfies
the condition
$$\text{\em (\c1{k-1})\ \&\ (\c2{k-1})\ \&\ (\c3{k-1}) };$$
\item[(ii)] there exists a point $V_k$ such that the $(k+1)$-gon $\P_{k+1}:=(V_0,\dots,V_{k-1},V_k)$ is quasi-strict and satisfies
the condition
$$\text{\em (\c1{k-1})\ \&\ (\c2{k-1})\ \&\ ($\nega$\c3{k-1}) };$$
here and in what follows, $\nega$ is the usual negation symbol, so that \break {\em ($\nega$\c3{k-1})} means that {\em (\c3{k-1})} does not hold; 
\item[(iii)] there exists a point $V_k$ such that the $(k+1)$-gon $\P_{k+1}:=(V_0,\dots,V_{k-1},V_k)$ is quasi-strict and satisfies
the condition
$$\text{\em (\c1{k-1})\ \&\ ($\nega$\c2{k-1})\ \&\ (\c3{k-1}) };$$ 
\item[(iv)] there exists a point $V_k$ such that the $(k+1)$-gon $\P_{k+1}:=(V_0,\dots,V_{k-1},V_k)$ is quasi-strict and satisfies
the condition
$$\text{\em ($\nega$\c1{k-1})\ \&\ (\c2{k-1})\ \&\ (\c3{k-1}) }.$$ 
\end{description}
\end{sublemma}

We shall prove this sublemma later. Now, let us complete the proof of Lemma~\ref{lem:minimality}. 

For each $\omega\in\{1,2,3\}$ and each set $J\subseteq\intr2{n-2}$, introduce the condition 
$$\text{(\c\omega J)}:=\bigl(\forall i\in J\ \text{(\c\omega i)}\bigr),$$
which is the conjunction of conditions (\c\omega i) over all $i\in J$. 

Consider the following statement, for $n\ge3$:
\begin{equation}\label{eq:min}
\parbox{4.2in}
{
\noindent
for every $i\in\intr2{n-2}$ there exists a quasi-strict $n$-gon $\P_n:=(V_0,\dots,V_{n-1})$ 
satisfying the condition \\
\quad (\c1{\intr2{n-2}})\ \&\ (\c2{\intr2{n-2}})\ \&\ (\c3{\intr2{n-2}\setminus\{i\} })\ \&\ ($\nega$\c3 i).
}
\tag{\M3 n}
\end{equation}

We shall prove statement \eqref{eq:min} by induction in $n$.
If $n=3$, then $\intr2{n-2}=\emptyset$, so that \eqref{eq:min} trivially holds.

Suppose next that statement \eqref{eq:min} holds for some $n=k$, where $k\ge3$. We have to verify that then statement \eqref{eq:min} holds for $n=k+1$. For $n=k+1$ and $i\in\intr2{n-2}$, only two cases are possible: $i\in\intr2{k-2}$ or $i=k-1$. Let us consider these two cases separately.

{\em Case 1\/}:\ $i\in\intr2{k-2}$.\quad In this case, by induction, there exists a quasi-strict $k$-gon $\P_k:=(V_0,\dots,V_{k-1})$ satisfying the condition
$$\text{(\c1{\intr2{k-2}})\ \&\ (\c2{\intr2{k-2}})\ \&\ (\c3{\intr2{k-2}\setminus\{i\} })\ \&\ ($\nega$\c3 i).}$$
By part (i) of Sublemma \ref{lem:minimal}, there exists a point $V_k$ such that 
the $(k+1)$-gon $\P_{k+1}:=(V_0,\dots,V_{k-1},V_k)$ is quasi-strict and satisfies
the condition
$$\text{(\c1{k-1})\ \&\ (\c2{k-1})\ \&\ (\c3{k-1}) }.$$
It follows that $\P_{k+1}$ satisfies the condition
\begin{equation}\label{eq:M3(k+1)i}
\text{(\c1{\intr2{k-1}})\ \&\ (\c2{\intr2{k-1}})\ \&\ (\c3{\intr2{k-1}\setminus\{i\} })\ \&\ ($\nega$\c3 i).}
\end{equation}

{\em Case 2\/}:\ $i=k-1$.\quad For every $k\ge3$, there is a quasi-strict $k$-gon $\P_k:=(V_0,\dots,V_{k-1})$ satisfying the condition 
$$\text{(\c1{\intr2{k-2}})\ \&\ (\c2{\intr2{k-2}})\ \&\ (\c3{\intr2{k-2}}).}$$
(This follows by induction using part (i) of Sublemma \ref{lem:minimal}.)
Let $\P_k$ be such a $k$-gon. 
By part (ii) of Sublemma \ref{lem:minimal}, there exists a point $V_k$ such that the $(k+1)$-gon $\P_{k+1}:=(V_0,\dots,V_{k-1},V_k)$ is quasi-strict and satisfies
the condition
$$\text{(\c1{k-1})\ \&\ (\c2{k-1})\ \&\ ($\nega$\c3{k-1}) },$$
so that \eqref{eq:M3(k+1)i} again holds---with $i=k-1$. 

Thus, statement \eqref{eq:min} takes place for $n=k+1$, and hence for all $n\ge3$. 
This implies that none of the $n-3$ conditions (\c3 i) with $i\in\intr2{n-2}$ in Lemma \ref{lem:test} can be omitted (because, by Lemma \ref{lem:def}, all of the conditions (\c3{i}) with $i\in\intr2{n-2}$ are necessary for polygon $\P$ to be quasi-strictly convex). 

Similarly (but using parts (iii) and (iv) of Sublemma \ref{lem:minimal} rather than part (ii) of it), one can show that none of the conditions (\c2 i) or 
(\c1 i) (with $i\in\intr2{n-2}$) in Lemma \ref{lem:test} can be omitted.  
\end{proof}

\begin{proof}[Proof of Sublemma \ref{lem:minimal}]
Since polygon $\P_k=(V_0,\dots,V_{k-1})$ is quasi-strict and $k\ge3$, the points $V_0$, $V_1$, and $V_{k-1}$ are non-collinear, so that w.l.o.g.\ 
$$V_0=(0,0),\quad V_1=(1,0),\quad V_{k-1}=(0,1).$$
Let also
$$V_{k-2}=(x,y),\quad V_k=(u,v)$$
for some real $x$, $y$, $u$, $v$. 
At that, the values of $x$ and $y$ are given to us, while the values of $u$ and $v$ we are free to choose. Note that $x\ne0$, because polygon $\P_k=(V_0,\dots,V_{k-1})$ is quasi-strict and hence the points $V_0$, $V_{k-2}$, and $V_{k-1}$ are non-collinear.  

Now, in view of Lemma \ref{lem:calculation}, conditions (\c1{k-1}), (\c2{k-1}), (\c3{k-1}) can be rewritten, respectively, as
\begin{align} 
(x-u+uy-vx)x & >0, \label{eq:*} \\
(x-u+uy-vx)(-u) & >0, \label{eq:***} \\
v & >0. \label{eq:**} 
\end{align} 
Now we are ready to prove parts (i)--(iv) of Sublemma \ref{lem:minimal}.

{\bf (i):}\quad For any given values of $x\ne0$ and $y$, let
$$u:=-\vp x,\quad v:=\vp+\vp^2$$
for $\vp\in(0,\vp_0)$, where $\vp_0:=0.1/(1+|y|)$. 
Then 
$$|-u+uy-vx|\le|u|(1+|y|)+|v||x|\le0.1|x|+0.11|x|<|x|,$$
whence
$\sign(x-u+uy-vx)=\sign x$, and so, all of the conditions \eqref{eq:*}, \eqref{eq:***}, \eqref{eq:**} hold; that is, conditions (\c1{k-1}), (\c2{k-1}), (\c3{k-1}) hold for all $V_k$ lying on the parabolic arc
$$P:=\{(-\vp x,\vp+\vp^2)\colon\vp\in(0,\vp_0)\}.$$
On the other hand, by Lemma \ref{lem:ordinary}, the $k$-gon $\P_k=(V_0,\dots,V_{k-1})$ is ordinary. Hence, for the $(k+1)$-gon $\P_{k+1}=(V_0,\dots,V_{k-1},V_k)$ not to be quasi-strict, the vertex $V_k$ must lie on the line through points $V_i$ and $V_j$ for some $i$ and $j$ such that $0\le i<j\le k-1$. But any one of these (finitely many) lines can have at most two points in common with the parabolic arc $P$; hence, 
the union of all such lines through points $V_i$ and $V_j$ cannot cover the infinite set $P$. This means that one can find a point $V_k$ in $P$ such that the $(k+1)$-gon $\P_{k+1}=(V_0,\dots,V_{k-1},V_k)$ is quasi-strict and satisfies conditions (\c1{k-1}), (\c2{k-1}), (\c3{k-1}). 

{\bf (ii):}\quad Similarly to the above, it can be seen that one can choose $V_k$ on the parabolic arc 
$$\{(-\vp x,-\vp-\vp^2)\colon\vp\in(0,\vp_0)\}$$
so that the $(k+1)$-gon $\P_{k+1}$ is quasi-strict and satisfies conditions (\c1{k-1}) and (\c2{k-1}) but not (\c3{k-1}). 

{\bf (iii):}\quad Similarly, one can choose $V_k$ on the parabolic arc 
$$\{(\vp x,\vp+\vp^2)\colon\vp\in(0,\vp_0)\}$$
so that the $(k+1)$-gon $\P_{k+1}$ is quasi-strict and satisfies conditions (\c1{k-1}) and (\c3{k-1}) but not (\c2{k-1}). 

{\bf (iv):}\quad Similarly, one can choose $V_k$ on the parabolic arc $$\{((1+\vp)x,(1+\vp)|y|+\vp^2)\colon\vp>0\}$$
so that the $(k+1)$-gon $\P_{k+1}$ is quasi-strict and satisfies
conditions (\c2{k-1}) and (\c3{k-1}) but not (\c1{k-1}). \big(Note that the conditions $u=(1+\vp)x$, $v=(1+\vp)|y|+\vp^2$, and $\vp>0$ imply
$$\frac1x(x-u+uy-vx)=-(\vp+\vp^2+(1+\vp)(|y|-y))<0.\ \big)$$
\end{proof}

\begin{proof}[Proof of Lemma \ref{lem:elim}]
Since one can do a cyclic permutation, it suffices to show that, if a polygon $\P_n=(V_0,\dots,V_{n-1})$ is quasi-strictly convex, then $\P_{n-1}=(V_0,\dots,V_{n-2})$ is so.

Observe that, if $n\le4$ and polygon $\P_n$ is quasi-strict, then $\P_{n-1}$ is quasi-strict. 
(Indeed, if $i\in\intr0{n-3}$ and $j\in\intr0{n-2}\setminus\{i,i+1\}$, then the points $V_i$, $V_{i+1}$, and $V_j$ are non-collinear, because polygon $\P_n=(V_0,\dots,V_{n-1})$ is quasi-strict. If $j\in\intr0{n-2}\setminus\{0,n-2\}$ and $n\le4$, then one must have $n=4$ and $j=1$, whence the points $V_{n-2}$, $V_0$, and $V_j=V_1$ are non-collinear, because polygon $\P_n=(V_0,\dots,V_{n-1})$ is quasi-strict.

Moreover, for $n\le4$ the $(n-1)$-gon $\P_{n-1}$ is always convex. Being also quasi-strict, $\P_{n-1}$ is then quasi-strictly convex. 

Assume now that $n\ge5$ and polygon $\P_n$ is quasi-strictly convex. Then, by Lemma \ref{lem:test}, one has (\c1 i), (\c2 i), and (\c3 i) for all $i\in\intr2{n-2}$ and hence for all $i\in\intr2{(n-1)-2}$. Therefore, Lemma \ref{lem:test} implies that $\P_{n-1}$ is quasi-strictly convex. 
\end{proof}

\begin{proof}[Proof of Lemma \ref{lem:strict}]
\textbf{``Only if"}\quad The ``only if" part of Lemma \ref{lem:strict} is trivial.

\textbf{``If"}\quad This part is proved by induction in $n$. The case $n\le2$ is trivial, because then there are no three distinct $i$, $j$, and $k$ in the set $\intr0{n-1}$. 

Let then $n\ge3$. Assume that the vertices $V_i$, $V_j$, and $V_k$ of polygon $\P_n:=(V_0,\dots,V_{n-1})$ are collinear for some distinct $i$, $j$, and $k$ in $\intr0{n-1}$. W.l.o.g., 
$0=i<j<k\le n-1$. Moreover, then $k\ne n-1$, because vertices $V_{n-1}$ and $V_n=V_0$ of polygon $\P_n$ are adjacent to each other. Hence, $0=i<j<k\le n-2$, so that the points $V_i$, $V_j$, and $V_k$ are vertices of polygon $\P_{n-1}:=(V_0,\dots,V_{n-2})$. But, by Lemma \ref{lem:elim}, polygon $\P_{n-1}$ is quasi-strictly convex. Hence, by induction, $V_i$, $V_j$, and $V_k$ are non-collinear.     
\end{proof}

\renewcommand{\refname}{\textsf{\bf Literature}}

\bigskip

{\parskip0pt \parindent0pt \it Department of Mathematical Sciences

Michigan Technological University

Houghton, MI 49931

USA

e-mail: ipinelis@mtu.edu}

\end{document}